\documentclass
[twocolumn,preprintnumbers,superscriptaddress,unsortedaddress,prl,tightenlines,footinbib]{revtex4}%
\usepackage{amssymb}
\usepackage{amsmath}
\usepackage{graphicx}
\usepackage{dcolumn}
\usepackage{bm}
\usepackage{amsfonts}%
\providecommand{\U}[1]{\protect\rule{.1in}{.1in}}
\begin{document}
\title{Intermode Dephasing in a Superconducting Stripline Resonator}
\author{Oren Suchoi}
\affiliation{Department of Electrical Engineering, Technion, Haifa 32000 Israel}
\author{Baleegh Abdo}
\affiliation{Department of Electrical Engineering, Technion, Haifa 32000 Israel}
\author{Eran Segev}
\affiliation{Department of Electrical Engineering, Technion, Haifa 32000 Israel}
\author{Oleg Shtempluck}
\affiliation{Department of Electrical Engineering, Technion, Haifa 32000 Israel}
\author{M. P. Blencowe}
\affiliation{Department of Physics and Astronomy, Dartmouth College, Hanover, New Hampshire
03755, USA}
\author{Eyal Buks}
\affiliation{Department of Electrical Engineering, Technion, Haifa 32000 Israel}
\date{\today }

\begin{abstract}
We study superconducting stripline resonator (SSR) made of Niobium, which is
integrated with a superconducting interference device (SQUID). The large
nonlinear inductance of the SQUID gives rise to strong Kerr nonlinearity in
the response of the SSR, which in turn results in strong coupling between
different modes of the SSR. We experimentally demonstrate that such intermode
coupling gives rise to dephasing of microwave photons. The dephasing rate
depends periodically on the external magnetic flux applied to the SQUID, where
the largest rate is obtained at half integer values (in units of the flux
quantum). To account for our result we compare our findings with theory and
find good agreement.

\end{abstract}

\maketitle





\section{Introduction}

Dephasing is the suppression process of quantum coherent effects due to
coupling between a quantum system and its external environment \cite{Zurek_36}%
. A Kerr nonlinearity in an electromagnetic resonator may lead to dispersive
intermode coupling, which in turn may result in dephasing of photons
\cite{Imoto_2287,Sanders_694}. Such a coupling mechanism can also be exploited
to allow quantum non demolition (QND) detection of single photons
\cite{Imoto_2287,Munro_033819,Santamore_144301,Buks_10001,Helmer_0712_1908}. A
Kerr nonlinearity exists in superconducting stripline resonators (SSR) due to
the effect of kinetic inductance. However, the resultant intermode coupling is
typically far too weak to allow any significant dephasing \cite{Buks_023815}.
On the other hand, a much stronger Kerr nonlinearity can be achieved by
integrating a superconducting interference device (SQUID) with the SSR
\cite{Yamamoto_042510,Sandberg_203501,Palacios-Laloy_1034}. External magnetic
flux can be employed in these devices to modulate both the linear and
nonlinear contributions to the inductance of the SQUID, which in turn allows
external control of both the resonance frequencies and the strength of Kerr
nonlinearity respectively. The enhanced Kerr nonlinearity also provides strong
coupling between different modes in the resonator that causes dephasing of one
mode (called the \textit{system} mode) when another one (the \textit{detector}
mode) is externally driven at relatively high powers.

Here, we employ such a configuration consisting of a Niobium SSR and
incorporating a SQUID device having a nanobridge in each of its two arms. We
monitor the resonance lineshape of one of the modes of the resonator (the
\textit{system} mode) as we simultaneously drive another one (the
\textit{detector} mode). We find that a significant broadening of the
resonance lineshape of the system mode occurs in the same region where the
response of the detector mode, which is measured simultaneously, becomes
strongly nonlinear. We provide theoretical evidence to substantiate our
hypothesis that the underlying mechanism responsible for the observed
broadening is intermode dephasing. The ability to externally control the
strength of the intermode coupling, which in turn controls the dephasing rate,
makes our device an ideal tool for experimentally studying fundamental issues
related to the quantum - classical transition \cite{Zurek_36}.%

\begin{figure}
[b]
\begin{center}
\includegraphics[
height=3.2846in,
width=3.0268in
]%
{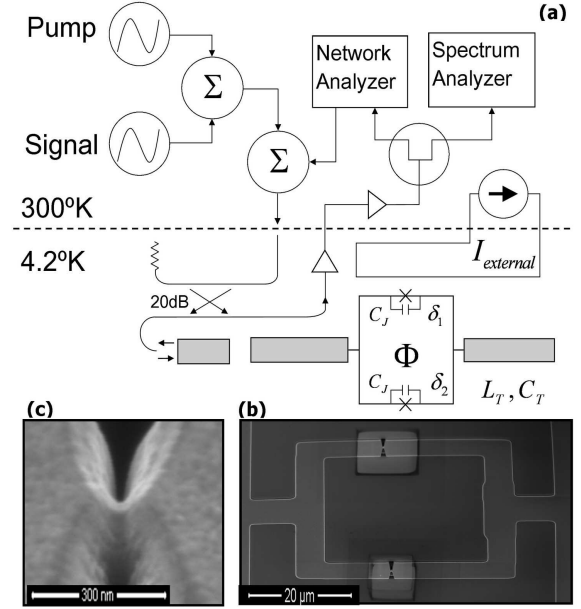}%
\caption{(a) The device and the experimental setup. The SSR is made of two
identical stripline sections of length $l_{\mathrm{T}}/2=110\operatorname{mm}%
$, each having inductance $L_{\mathrm{T}}$ and capacitance $C_{\mathrm{T}}$
per unit length and characteristic impedance $Z_{\mathrm{T}}=\sqrt
{L_{\mathrm{T}}/C_{\mathrm{T}}}=50\operatorname{\Omega }$. The stripline
sections are connected by a SQUID consisting of nanobridge type weak links.
All measurements are done at liquid Helium temperature $4.2\operatorname{K}$,
where the device is placed inside a copper package, which is internally coated
with Nb (to reduce surface resistance and to provide magnetic shielding). (b)
SEM micrograph of the SQUID (tilted view). The loop area is 39$\times
39\operatorname{\mu m}^{2}.$ (c) SEM micrograph of the nanobridge. }%
\label{setup}%
\end{center}
\end{figure}

\section{\bigskip Experimental System}

Fig. (\ref{setup}) schematically shows the device. The SSR \cite{Clark_3042,
Nation_104516, Palacios-Laloy_1034} comprises two identical stripline sections
connected by a SQUID. A nanobridge \cite{Likharev_101, Troeman_024509,
Lam_1078, Podd_134501} on each arm of the SQUID loop serves as a weak link.
The critical currents of the nanobridges are denoted by $I_{\mathrm{c1}}$ and
$I_{\mathrm{c2}}$ respectively. Both nanobridges are assumed to have the same
capacitance $C_{\mathrm{J}}$. The self inductance of the loop is denoted by
$\Lambda$. A feedline, which is weakly coupled to the SSR, is employed to
deliver the input and output microwave signals. The experimental setup is
presented in subplot (a) of Fig. (\ref{setup}).

The fabrication process starts with a high resistivity Si substrate coated
with SiN layers of thickness $100%
\operatorname{nm}%
$ on both sides. A $150%
\operatorname{nm}%
$ thick Nb layer is deposited on the wafer using magnetron DC sputtering.
Then, e-beam lithography and a subsequent liftoff process are employed to
pattern an Al mask, which defines the SSR and the SQUID leads. The device is
then etched using electron cyclotron resonance system with CF$_{4}$ plasma.
The nanobridges are fabricated using FEI Strata 400 Focus Ion Beam (FIB)
system \cite{Hao_192507,Hao_392,Bell_630, Datesman_928, Troeman_2152} at
accelerating voltage of $30%
\operatorname{kV}%
$ and Ga ions current of $9.7$ pA. The outer dimensions of the bridges are
about $150\times50%
\operatorname{nm}%
.$ However, the actual dimensions of the weak-links are smaller, since the
bombarding Ga ions penetrate into the Nb layer, and consequently, suppress
superconductivity over a depth estimated between $30%
\operatorname{nm}%
$ to $50%
\operatorname{nm}%
$ \cite{Troeman_2152, Datesman_3524}.

\section{Effective Hamiltonian}

The effective Hamiltonian of the closed system consisting of the SSR and the
SQUID , expressed in terms of the annihilation and creation operators $A_{1}$,
$A_{1}^{\dag},$ $A_{3}$ and $A_{3}^{\dag}$ of the first and third modes
respectively, is given by:
\begin{align}
\mathcal{H}_{\mathrm{eff}}  &  =\hbar\omega_{1}N_{1}+\hbar\omega_{3}%
N_{3}+V_{\mathrm{in}}\nonumber\\
&  +\hbar K_{1}N_{1}^{2}+\hbar\lambda_{1,3}N_{1}N_{3}\ . \label{H_eff}%
\end{align}
where $N_{1}=A_{1}^{\dag}A_{1}$ and $N_{3}=A_{3}^{\dag}A_{3}$ are number
operators, $V_{\mathrm{in}}=\hbar\sqrt{2\gamma_{\mathrm{f}1}}b_{1}%
^{\mathrm{in}}\left(  e^{-i\omega_{\mathrm{p}}t}A_{1}+e^{i\omega_{\mathrm{p}%
}t}A_{1}^{\dag}\right)  $ represents the external driving, $\gamma
_{\mathrm{f}1}$ is the coupling constant between the first mode and the
feedline, $b_{1}^{\mathrm{in}}$ is the amplitude of the driving pump tone
which is injected into the feedline to excite the first mode, and where
$\omega_{\mathrm{p}}$ is its angular frequency. Full Derivation of the
Hamiltonian is given in appendix A. The last two terms represent the Kerr
nonlinearity term of the first, externally driven (detector) mode and the
intermode coupling between the first and the third (system) modes,
respectively. The coefficients $\omega_{1},$ $\omega_{3},$ $K_{1}$ and
$\lambda_{1,3}$, which are calculated in appendix A., depend periodically on
the external flux $\Phi_{x}$ with period $\Phi_{0}$. The flux dependence of
$\omega_{1}$ and $\omega_{3}$ can be attributed to the inductance of the
SQUID, which is proportional to the second derivative of $\varepsilon_{0}$
with respect to $I,$ where $\varepsilon_{0}$ is the ground state energy of the
SQUID. On the other hand, both the Kerr nonlinearity $K_{1}$ and intermode
coupling $\lambda_{1,3}$ coefficients are proportional to the nonlinear
inductance of the SQUID \cite{Yurke_5054}, which in turn is proportional to
the fourth derivative of $\varepsilon_{0}$ with respect to $I$.

\section{\bigskip Resonance Frequency Shift}%

\begin{figure}
[b]
\begin{center}
\includegraphics[
height=5.0367in,
width=3.3096in
]%
{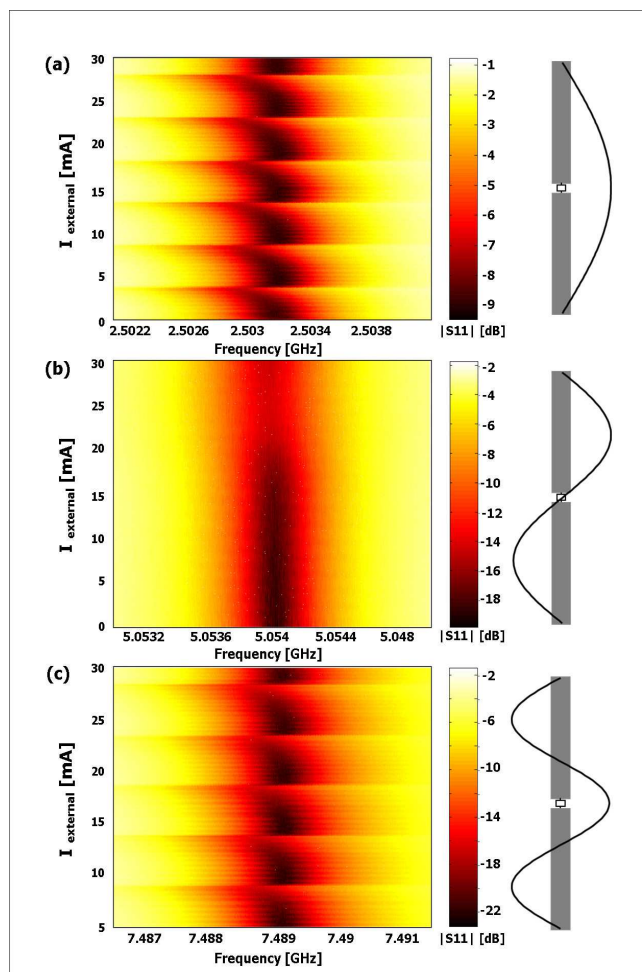}%
\caption{(Color online) Reflection coefficient $\left\vert S_{11}\right\vert $
vs. frequency and external flux for the first 3 modes of the SSR. A change of
$4.8\operatorname{mA}$ in the external current corresponds to a change of
$\Phi_{0}$ in the magnetic flux.}%
\label{3modes}%
\end{center}
\end{figure}
%

\begin{figure}
[b]
\begin{center}
\includegraphics[
height=2.6273in,
width=3.2396in
]%
{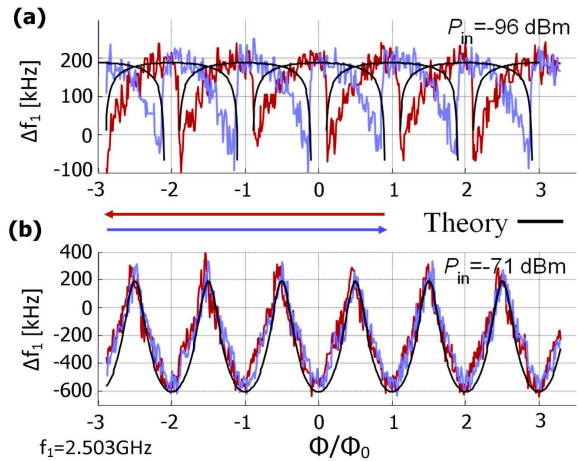}%
\caption{(Color online) The resonance frequency shift $\Delta f_{1}$ of the
first (detector) mode vs. applied flux for two different values of
$P_{\mathrm{in}}$. The flux is first swept upwards (blue line) and than
downwards (red line). The black solid lines represent the theoretical
calculation of $\Delta f_{1}$ using the following parameters: $\beta_{L}=7.4$
for $P_{\mathrm{in}}=-96$ dBm, $\beta_{L}=0.15$ for $P_{\mathrm{in}}=-71$ dBm,
and $I_{\mathrm{c1}}/I_{\mathrm{c2}}=3$ for both cases.}%
\label{f1}%
\end{center}
\end{figure}

Figure (\ref{3modes}) shows measurements of the reflection coefficient
$\left\vert S_{11}\right\vert $ ($S_{11}$ is the ratio between the reflected
outgoing and the injected incoming amplitudes in the feedline) of the first 3
modes of the resonator as a function of frequency and externally applied flux
$\Phi_{x}$. The sketches on the right hand side show the current waveform of
each mode. For the first and the third modes, $S_{11}$ is found to be a
periodic function of $\Phi_{x}$ with period $\Phi_{0}$, where $\Phi_{0}=h/2e$
is the flux quantum. On the other hand, the 2nd mode, which is decoupled from
the SQUID since its current waveform has a node at the location of the SQUID,
does not exhibit a flux dependence. Note that the data in Fig. (\ref{3modes})
is obtained by sweeping the magnetic flux upwards. However, as can be seen
from Fig. (\ref{f1}a), in which the resonance frequency $f_{1}=\omega_{1}%
/2\pi$ of the first mode is measured versus both increasing (blue) and
decreasing (red) magnetic flux, the response is hysteretic.

The solid black line in Fig. (\ref{f1}a) is obtained by numerically evaluating
the resonance frequency $f_{1}=\omega_{1}/2\pi$ using Eq.
(\ref{frequency_shift}) in appendix A. For the parameters that are used in the
calculation for this case (see figure caption), the SQUID can be either
monostable or bistable depending on $\Phi_{x}$. Consequently, sharp
transitions occur near the values of $\Phi_{x}$ corresponding to a boundary
between these regions, and as a result, the response is hysteretic.
Interestingly, as the input power is increased the response becomes
non-hysteretic, as can be seen from Fig. (\ref{f1}b), which shows a
measurement of $f_{1}$ at $P_{\mathrm{in}}=-71$ dBm. Theoretically, this
behavior is accounted for by assuming that the value of the screening
parameter $\beta_{L}=2\pi\Lambda I_{\mathrm{c}}/\Phi_{0}$, which is
proportional to the average critical current $I_{\mathrm{c}}=\left(
I_{\mathrm{c1}}+I_{\mathrm{c2}}\right)  /2$, is significantly lower for this
case ($0.15$ instead of the value $7.4$, which was used to fit the data for
$P_{\mathrm{in}}=-96$ dBm). To account for this behavior we discuss in
appendix B. the possibility that local heating of the nanobridges is
responsible for the drop in $I_{\mathrm{c}}$ at elevated input powers.
Assuming that the heat is mainly dissipated down into the substrate rather
than along the film, we estimate that the temperature rise for $P_{\mathrm{in}%
}=-70$ dBm is $4%
\operatorname{K}%
$. This rough estimation indicates that heating may indeed play an important
role, and may be held responsible for the apparent drop in the critical current.

\section{Intermodulation and Intermode Dephasing}

%

\begin{figure}
[b]
\begin{center}
\includegraphics[
trim=0.190219in 0.000000in 0.000000in 0.035603in,
height=2.3436in,
width=3.2872in
]%
{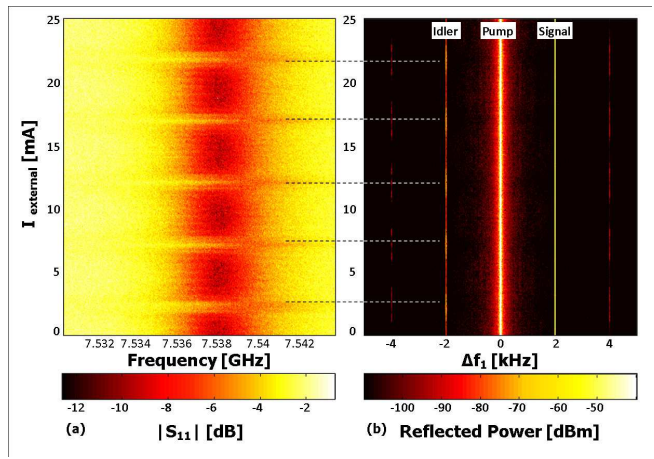}%
\caption{(Color online) IM characterization of the detector mode [panel (b)]
and $\left\vert S_{11}\right\vert $ measurements of the system mode [panel
(a)]. Largest idler gain as well as highest dephasing rate is obtained at half
integer values of the externally applied flux.}%
\label{IntM_Dep}%
\end{center}
\end{figure}

\begin{figure}
[t]
\begin{center}
\includegraphics[
height=2.5425in,
width=3.3399in
]%
{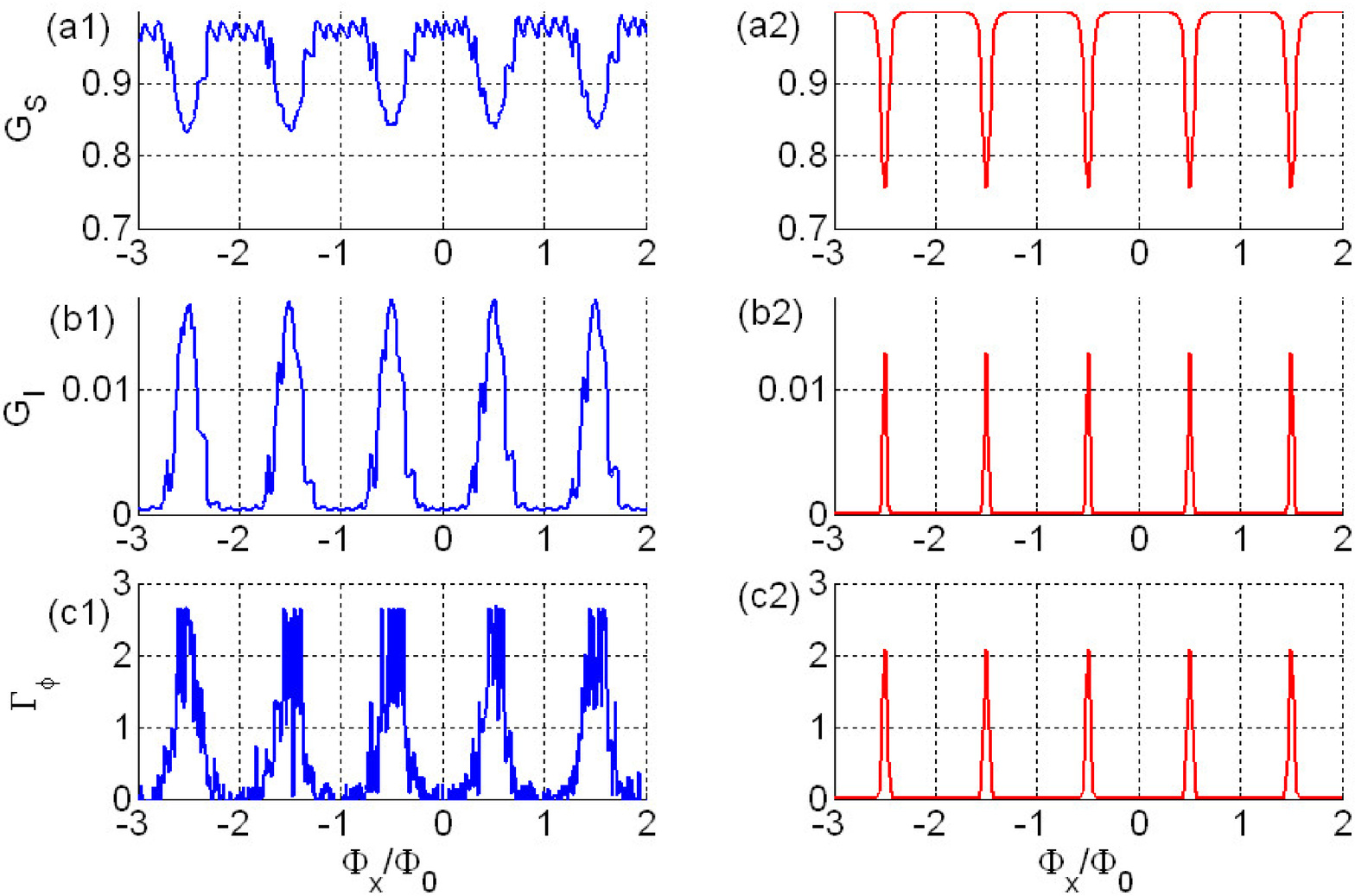}%
\caption{(Color online) Signal gain $G_{\mathrm{s}}$, idler gain
$G_{\mathrm{i}}$, and normalized dephasing rate $\Gamma_{\varphi}$.
Experimental results are shown in panels (a1), (b1) and (c1), whereas
theoretical predictions are shown in panels (a2), (b2) and (c2). The following
device parameters were taken in order to evaluate $G_{\mathrm{s}}$ and
$G_{\mathrm{i}}$ theoretically using Eqs. (82) and (83) of Ref.
\cite{Yurke_5054} [panels (a2) and (b2) respectively)], and to evaluate
$\Gamma_{\varphi}$ using Eq. (70) of Ref. \cite{Buks_023815} [panel (c2)]:
$\left(  \gamma_{\mathrm{f}3}+\gamma_{\mathrm{d}3}\right)  /\omega_{1}=5000$,
$\gamma_{\mathrm{f}3}/\gamma_{\mathrm{d}3}=0.15$, $I_{\mathrm{c1}%
}=1.5\operatorname{\mu A}$ and $I_{\mathrm{c2}}=4.5\operatorname{\mu A}$. Note
that the effect of nonlinear damping \cite{Yurke_5054} is disregarded.}%
\label{GS_GI_DR}%
\end{center}
\end{figure}

Both nonlinear terms in the Hamiltonian $\mathcal{H}_{\mathrm{eff}}$
(\ref{H_eff}) play an important role as $P_{\mathrm{in}}$ is increased. The
effect of the Kerr nonlinearity can be sensitively observed by employing
intermodulation (IM) characterization \cite{Yurke_5054}. In this method, in
addition to the relatively strong \textit{pump} tone at frequency
$\omega_{\mathrm{p}}$, which is used to drive the first mode to any desirable
operating point, another tone, called \textit{signal}, which has a much
smaller power $P_{\mathrm{s,in}}$ and a nearby frequency $\omega_{\mathrm{p}%
}+\delta\omega$ ($\delta\omega$ is much smaller than the resonance width), is
also injected simultaneously into the feedline. Due to Kerr nonlinearity these
two inputs may mix in the resonator and produce tones of IM products.
Typically, the largest IM products are the output signal at frequency
$\omega_{\mathrm{p}}+\delta\omega$ and the output idler at frequency
$\omega_{\mathrm{p}}-\delta\omega$. The two corresponding gain factors, namely
the signal gain $G_{\mathrm{s}}=P_{\mathrm{s,out}}/P_{\mathrm{s,in}}$ and the
idler gain $G_{\mathrm{i}}=P_{\mathrm{i,out}}/P_{\mathrm{s,in}}$, where
$P_{\mathrm{s,out}}$ and $P_{\mathrm{i,out}}$ are the powers of the output
signal and output idler tones respectively, were evaluated in Ref.
\cite{Yurke_5054}. Panel (b) of Fig. (\ref{IntM_Dep}) presents a color map
showing IM characterization of the first (detector) mode, which was obtained
using a spectrum analyzer. The powers of the injected pump and signal tones in
the IM measurement are $-62.1$ dBm and $-81$ dBm respectively. Both
$G_{\mathrm{s}}$ and $G_{\mathrm{i}}$ periodically oscillate as a function of
the external current. This behavior is seen more clearly in panels (a1) and
(b1) of Fig. (\ref{GS_GI_DR}) which exhibit $G_{\mathrm{s}}$ and
$G_{\mathrm{i}}$ versus $\Phi_{x}/\Phi_{0}$.%

The intermode coupling term in the Hamiltonian (\ref{H_eff}) can be exploited
to continuously measure the number of photons in the system mode by externally
driving the detector mode \cite{Imoto_2287,Sanders_694}. Such a measurement
scheme is characterized by the time it takes to resolve adjacent number states
of the system mode. Significant dephasing occurs when this time scale is made
comparable or shorter than the lifetime of photons in the system mode.
Theoretically, dephasing of photons in the system mode is expected to give
rise to a resonance frequency shift and to broadening of the resonance line
shape of the power reflection coefficient $\left\vert S_{11}\left(
\omega\right)  \right\vert ^{2}$, which is given by \cite{Levinson_299}%
\begin{equation}
\left\vert S_{11}\left(  \omega\right)  \right\vert ^{2}=1-\frac
{\gamma_{\mathrm{f}3}\gamma_{\mathrm{d}3}}{\gamma_{\mathrm{f}3}+\gamma
_{\mathrm{d}3}}\frac{4\gamma_{\mathrm{tot}}}{\gamma_{\mathrm{tot}}^{2}+\left(
\omega-\tilde{\omega}_{3}\right)  ^{2}}\;, \label{S_11}%
\end{equation}
where $\tilde{\omega}_{3}$ is the shifted angular resonance frequency, the
total width is given by $\gamma_{\mathrm{tot}}=\gamma_{\mathrm{f}3}%
+\gamma_{\mathrm{d}3}+1/\tau_{\varphi}$, where $\gamma_{\mathrm{f}3}$ denotes
the coupling constant between the system mode and the feedline, and
$\gamma_{\mathrm{d}3}$ denotes the damping rate of the system mode and
$1/\tau_{\phi}$ is the dephasing rate of photons in the system mode.

Simultaneously with the IM characterization, we also measure the resonance
line shape of the third mode using a network analyzer. A very low input power
of $-101$ dBm is employed to avoid any nonlinear response of the third mode.
As can be seen from the results, which are presented in panel (a) of Fig.
(\ref{IntM_Dep}), the measured reflection coefficient $\left\vert
S_{11}\left(  \omega\right)  \right\vert ^{2}$ periodically oscillates as a
function of $\Phi_{x}$. Fitting the experimental data to Eq. (\ref{S_11})
yields the normalized dephasing rate $\Gamma_{\varphi}=1/\left(
\gamma_{\mathrm{f}3}+\gamma_{\mathrm{d}3}\right)  \tau_{\varphi}$. As can be
seen from Fig. (\ref{GS_GI_DR}), at the same points where $G_{\mathrm{i}}$
peaks [panel (b1)], namely for half integer values of the external flux, a
strong peak is found in $\Gamma_{\varphi}$ [panel (c1)]. At these points, the
value of \ $\Gamma_{\varphi}$ exceeds unity, namely, the dephasing rate
becomes larger than the system mode decay rate.

To account for the experimental results, we employ Eqs. (82) and (83) of Ref.
\cite{Yurke_5054} to calculate the gain factors $G_{\mathrm{s}}$ and
$G_{\mathrm{i}}$ respectively, and Eq. (70) of Ref. \cite{Buks_023815} to
calculate the normalized dephasing rate $\Gamma_{\varphi}$. The results, given
in panels (a2), (b2) and (c2) of Fig. (\ref{GS_GI_DR}) yield fairly good
agreement with the experimental data [panels (a1), (b1) and (c1)]. The device
parameters that were used in the calculation are listed in the figure caption.
The flux dependence of the gain factors $G_{\mathrm{s}}$ and $G_{\mathrm{i}}$
and that of the normalized dephasing rate $\Gamma_{\varphi}$ can be attributed
to the periodic flux dependence of the parameters $\omega_{1}$, $\omega_{3}$,
$K_{1}$ and $\lambda_{1,3}$ of the Hamiltonian (\ref{H_eff}). Both nonlinear
parameters $K_{1}$ and $\lambda_{1,3}$ peak at half integer values of the
external flux. Consequently, both $G_{\mathrm{i}}$, which can be considered as
a measure of the strength of nonlinearity, and $\Gamma_{\varphi}$, which
strongly depends on $\lambda_{1,3}$, obtain their largest values at these points.

\section{Conclusion}

Integrating a SQUID having large nonlinear inductance with an SSR leads to
strong IM distortion and strong intermode coupling. In the present paper we
have exploited these effects to study a novel mechanism of dephasing of
microwave photons that can be externally controlled. The same intermode
coupling that is responsible for the observed photon dephasing can also be
exploited for single photon detection\cite{Imoto_2287,Sanders_694}. In future
experiments several improvements, such as increasing the nonlinear coupling,
as well as reducing the temperature and using lower noise pre-amplifier should
allow detection of single microwave photons.

\section{Acnowledgments}

This work was supported by the Israel Science Foundation, USA-Israel
Binational Science Foundation, Germany Israel Foundation, the Deborah
Foundation, the Poznanski Foundation, Russel Berrie nanotechnology institute,
and MAFAT. BA was supported by the Ministry of Science, Culture and Sports.

\section{Appendix A: Detailed derivation of the effective Hamiltonian}

The effective Hamiltonian of the closed system comprising the SSR and the
SQUID \cite{Clark_3042, Blencowe_014511} is found using the same method that
was previously employed in Refs. \cite{Blencowe_014511,Nation_104516}. Here
however, we relax the assumption that the self inductance of the SQUID loop is
small, and also the assumption that both junctions have the same critical
currents. On the other hand, we assume that the inductance of the SQUID, which
is denoted as $L_{\mathrm{S}}$, is much smaller than the total inductance of
the stripline $L_{\mathrm{T}}l_{\mathrm{T}}$. This assumption can be justified
by considering the fact that the measured angular resonance frequencies
$\omega_{n}$ of the first 3 modes ($n\in\left\{  1,2,3\right\}  $) for all
values of $\Phi_{x}$ (see Figs. 2 and 3 in the paper body) are very close to
the values expected from a uniform resonator having length $l_{\mathrm{T}}$,
namely $n\omega_{\mathrm{T}}$, where $\omega_{\mathrm{T}}=\pi/l_{\mathrm{T}%
}\sqrt{L_{\mathrm{T}}C_{\mathrm{T}}}$. Moreover, the normalized flux-induced
shift $\Delta\omega_{n}/n\omega_{\mathrm{T}}$ in the angular resonance
frequency of the first 3 modes is quite small and never exceeds $10^{-3}$.
Both observations indicate that the ratio $L_{\mathrm{S}}/L_{\mathrm{T}%
}l_{\mathrm{T}}$ can indeed be considered as a small parameter.

The resultant Hamiltonian of the closed system is given by $\mathcal{H}%
=\mathcal{H}_{\mathrm{SSR}}+\mathcal{H}_{\mathrm{S}}\left(  I\right)  $, where
$\mathcal{H}_{\mathrm{SSR}}$ is the SSR Hamiltonian and where $\mathcal{H}%
_{\mathrm{S}}\left(  I\right)  $ is the SQUID Hamiltonian, which depends on
the current $I$ at the center of the SSR, namely, the current flowing through
the SQUID. In terms of annihilation ($A_{1}$ and $A_{3}$) and creation
($A_{1}^{\dag}$ and $A_{3}^{\dag}$) operators for the first and third modes of
the SSR respectively, the Hamiltonian $\mathcal{H}_{\mathrm{SSR}}$ can be
expressed as
\begin{equation}
\mathcal{H}_{\mathrm{SSR}}=\hbar\omega_{\mathrm{T}}\left(  N_{1}%
+3N_{3}\right)  +V_{\mathrm{in}}\;,
\end{equation}
where $N_{1}=A_{1}^{\dag}A_{1}$ and $N_{3}=A_{3}^{\dag}A_{3}$ are number
operators,%
\begin{equation}
V_{\mathrm{in}}=\hbar\sqrt{2\gamma_{\mathrm{f}1}}b_{1}^{\mathrm{in}}\left(
e^{-i\omega_{\mathrm{p}}t}A_{1}+e^{i\omega_{\mathrm{p}}t}A_{1}^{\dag}\right)
\end{equation}
represents the external driving, $\gamma_{\mathrm{f}1}$ is the coupling
constant between the 1st mode and the feedline, $b_{1}^{\mathrm{in}}$ is the
amplitude of the driving pump tone, which is injected into the feedline to
excite the first mode, and $\omega_{\mathrm{p}}$ is its angular frequency.

\subsection{The kinetic inductance of the nanobridges}

The Hamiltonian for the SQUID depends on the properties of the nanobridges.
Due to the Ga ions implanted in the outer layer of the Niobium during the FIB
process and the consequent suppression of superconductivity in that layer
\cite{Troeman_2152, Datesman_3524}, the weak links are treated as variable
thickness nanobridges. The behavior of such a nanobridge is strongly dependent
on the ratio $l/\xi$
\cite{Granata_275501,Hasselbach_4432,Hasselbach_140,Baratoff_1096,Likharev_101,Likharev_950,
Gumann_064529, Podd_134501}, where $l$ is the bridge length and $\xi$ is the
coherence length of the Cooper pairs. The coherence length $\xi$ depends also
on the temperature of the bridge. In the dirty limit $\xi$ is given by
$\xi(T)=0.852\sqrt{\xi_{0}l_{f}\left(  T_{C}/T-1\right)  ^{-1}}$
\cite{Likharev_101}, where $\xi_{0}$ is the size of the cooper pair and
$l_{f}$ is the mean free path\cite{Pronin_14416,Maxfield_A1515}. The
current-phase relation (CPR) of the bridges is periodic with respect to the
gauge invariant phase $\delta$ across the bridge. When $l/\xi(T)\ll1,$\ the
nanobridge behaves like a regular Josephson junction (JJ) with a sinusoidal
CPR\cite{Golubov_411}. However, as the ratio $l/\xi(T)$ becomes larger, the
CPR deviates from the sinosoidal form and can also become multivalued
\cite{Likharev_101}. In case the CPR is not multivalued the bridge can be
approximately considered as a JJ having an extra kinetic inductance
$L_{\mathrm{K}}$ . The effect of the kinetic inductance can be taken into
account by replacing the screening parameter of the loop $\beta_{L}%
=2\pi\Lambda I_{c}/\Phi_{0}$ by an effective one given by $\beta_{L}%
+\Delta\beta$, where $\Delta\beta=2\pi L_{\mathrm{K}}I_{\mathrm{c}}/\Phi_{0}$.

\bigskip In order to estimate $\Delta\beta$ we use Eqs. (47)-(49) and the data
in Fig. 5 of Ref. \cite{Troeman_024509}. For $l/\xi=1.7$ the bridges'
contribution is $\Delta\beta\simeq1$ . As we will discuss below, both
$\beta_{L}$ and $\Delta\beta$ depend on the injected power $P_{\mathrm{in}}$
that is used to excite the resonator due to a heating effect. However, for all
values of $P_{\mathrm{in}}$ that were used in our experiment, we estimate that
the ratio $\Delta\beta/\beta_{L}$ never exceeds the value $0.5$ and thus the
effect of kinetic inductance can be considered as small. Furthermore, the CPR
remains a single valued function in the entire range of parameters that is
explored in our experiments. Consequently, the nanobridges can be treated as
regular JJs to a good approximation.

\subsection{The SQUID Hamiltonian}

In the following derivation we treat the nanobridges as regular JJs. We
consider the case where the critical currents of both nanobridges are
$I_{\mathrm{c1}}=I_{\mathrm{c}}\left(  1+\alpha\right)  $ and $I_{\mathrm{c2}%
}=I_{\mathrm{c}}\left(  1-\alpha\right)  $ respectively, where the
dimensionless parameter $\alpha$ characterizes the asymmetry in the SQUID. The
Hamiltonian for the SQUID, which is expressed in terms of the two gauge
invariant phases $\delta_{1}$ and $\delta_{2}$ across both junctions, and
their canonical conjugates $p_{1}$ and $p_{2}$, is given by%
\begin{equation}
\mathcal{H}_{\mathrm{S}}\left(  I\right)  =\frac{2\pi\omega_{p}^{2}\left(
p_{1}^{2}+p_{2}^{2}\right)  }{E_{0}}+E_{0}u\left(  \delta_{1},\delta
_{2};I\right)  \;,
\end{equation}
where $\omega_{pl}=\sqrt{I_{c}/C_{\mathrm{J}}\Phi_{0}}$ is the plasma
frequency, $E_{0}=\Phi_{0}I_{c}/\pi$ is the Josephson energy, and the
dimensionless potential $u$ is given by \cite{Mitra_214512}%
\begin{multline}
u=-\frac{\left(  1+\alpha\right)  \cos\delta_{1}+\left(  1-\alpha\right)
\cos\delta_{2}}{2}+\frac{\left(  \frac{\delta_{1}-\delta_{2}}{2}+\frac{\pi
\Phi_{x}}{\Phi_{0}}\right)  ^{2}}{\beta_{L}}\label{u=}\\
-\frac{\left(  \delta_{1}+\delta_{2}\right)  I}{4I_{c}}-\frac{\zeta\left(
\delta_{1}+\delta_{2}\right)  ^{2}}{16}\;,
\end{multline}
where $\zeta=\Phi_{0}/2I_{c}L_{\mathrm{T}}l_{\mathrm{T}}$.

\subsection{Adiabatic approximation}

Due to the extremely small capacitance $C_{\mathrm{J}}$ of both nanobridges
\cite{Ralph_10753}, the plasma frequency $\omega_{\mathrm{pl}}$ of the SQUID
is estimated to exceed $1%
\operatorname{THz}%
$. Thus, the effect of the SQUID on the SSR, which has a much slower dynamics,
can be treated using the adiabatic approximation \cite{Buks_10001,Buks_026217}%
. Formally, treating the current $I$ as a parameter (rather than a degree of
freedom), the Hamiltonian $\mathcal{H}_{\mathrm{S}}$ can be diagonalized
$\mathcal{H}_{\mathrm{S}}\left\vert k\left(  I\right)  \right\rangle
=\varepsilon_{k}\left(  I\right)  \left\vert k\left(  I\right)  \right\rangle
$, where $k=0,1,2,...$, and $\left\langle k\left(  I\right)  |l\left(
I\right)  \right\rangle =\delta_{kl}$. To lowest order in the adiabatic
expansion the effective Hamiltonian governing the dynamics of the slow degrees
of freedom corresponding to the fast part of the system occupying the state
$\left\vert k\left(  I\right)  \right\rangle $ is given by $\mathcal{H}%
_{k}^{\mathrm{A}}=\mathcal{H}_{\mathrm{SSR}}+\varepsilon_{k}\left(  I\right)
$ \cite{Littlejohn_5239,Panati_250405}. Furthermore, in the limit where the
thermal energy $k_{\mathrm{B}}T$ is much smaller than the typical energy
spacing between different levels of $\mathcal{H}_{1}$ ($\simeq\hbar
\omega_{\mathrm{pl}}$) one can assume that the SQUID remains in its current
dependent ground state $\left\vert 0\left(  I\right)  \right\rangle $. For
most cases this assumption is valid for our experimental parameters. It is
important, however, to note that when the externally applied magnetic flux is
close to a half-integer value (in units of $\Phi_{0}$), namely, when $\Phi
_{x}\simeq\left(  n+1/2\right)  \Phi_{0}$, where $n$ is integer, this
approximation may break down. Near these points the potential $u$ may have two
different neighboring wells having similar depth. Consequently, near these
points, the energy gap between the ground state and the first excited state
can become much smaller than $\hbar\omega_{\mathrm{pl}}$. On the other hand,
the ratio between the height of the barrier separating the two wells ($\simeq
E_{0}$) and the energy spacing between intra-well states ($\simeq\hbar
\omega_{\mathrm{pl}}$) is typically $E_{0}/\hbar\omega_{\mathrm{pl}}\simeq100$
for our samples. Since the coupling between states localized in different
wells depends exponentially on this ratio, we conclude that to a good
approximation the inter-well coupling can be neglected. Moreover, in the same
limit where $E_{0}/\hbar\omega_{\mathrm{pl}}\gg1$, one can approximate the
ground state energy $\varepsilon_{0}$ by the value of $E_{0}u$ at the bottom
of the well where the system is localized.

The current $I$ at the center of the SSR can readably be expressed in terms of
the annihilation and creation operators $A_{1}$, $A_{1}^{\dag}$ $A_{3}$ and
$A_{3}^{\dag}$. This allows expanding the current dependent ground state
energy $\varepsilon_{0}\left(  I\right)  $ as a power series of these
operators. In the rotating wave approximation oscillating terms in such an
expansion are neglected since their effect on the dynamics for a time scale
much longer than a typical oscillation period is negligibly small. Moreover,
constant terms in the Hamiltonian are disregarded since they only give rise to
a global phase factor. In the present experiment the 1st SSR mode is
externally driven, and we focus on the resultant dephasing induced on the 3rd
mode. To that end we include in the effective Hamiltonian of the closed system
in addition to the linear terms corresponding to the 1st and 3rd modes, also
the Kerr nonlinearity term of the 1st mode, which is externally driven, and
also the term representing intermode coupling between the 1st and the 3rd
modes [see Eq. (\ref{H_eff})].

The angular resonance frequency shift of the 1st and the 3rd modes, which is
given by%
\begin{equation}
\frac{\omega_{1}-\omega_{\mathrm{T}}}{\omega_{\mathrm{T}}}=\frac{\omega
_{3}-3\omega_{\mathrm{T}}}{3\omega_{\mathrm{T}}}=\zeta\frac{\partial
^{2}\left(  \varepsilon_{0}/E_{0}\right)  }{\partial\left(  I/I_{\mathrm{c}%
}\right)  ^{2}}\;, \label{frequency_shift}%
\end{equation}
can be attributed to the inductance of the SQUID, which is proportional to the
second derivative of $\varepsilon_{0}$ with respect to $I$. On the other hand,
the Kerr nonlinearity, which is given by%
\begin{equation}
\frac{K_{1}}{\omega_{1}}=\frac{\zeta^{2}\hbar\omega_{1}}{2E_{0}}\frac
{\partial^{4}\left(  \varepsilon_{0}/E_{0}\right)  }{\partial\left(
I/I_{\mathrm{c}}\right)  ^{4}}\;,
\end{equation}
and the intermode coupling, which is given by $\lambda_{1,3}=9K_{1}$, can both
be attributed to the nonlinear inductance of the SQUID \cite{Yurke_5054},
which is proportional to the fourth derivative of $\varepsilon_{0}$ with
respect to $I$.

\subsection{Evaluation of $\omega_{1},$ $\omega_{3},$ $K_{1}$ and
$\lambda_{1,3}$ in the limit $\beta_{L}\ll1$}%

\begin{figure}
[b]
\begin{center}
\includegraphics[
height=2.0003in,
width=3in
]%
{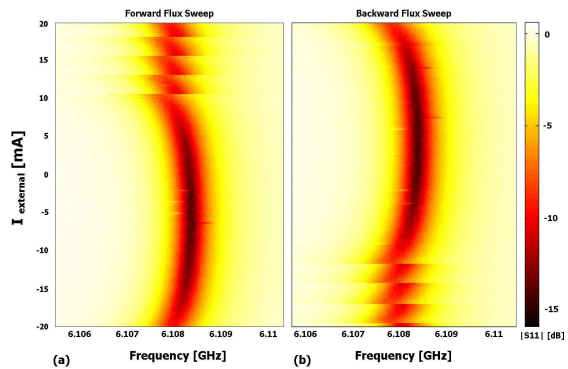}%
\caption{Measured $\left\vert S_{11}\right\vert $ at input power
$P_{\mathrm{in}}=-95$dBm for forward (a) and backward (b) magnetic flux sweep.
In this sample $\beta_{L}=20,$ and the response is highly hysteretic.}%
\label{largebetal}%
\end{center}
\end{figure}

The evaluation of the parameters $\omega_{1}$, $\omega_{3}$, $K_{1}$ and
$\lambda_{1,3}$ generally requires a numerical calculation. However, an
analytical approximation can be employed when $\beta_{L}\ll1$. In this limit
the phase difference $\delta_{2}-\delta_{1}$ is strongly confined near the
value $2\pi\Phi_{x}/\Phi_{0}$, as can be seen from Eq. (\ref{u=}). This fact
can be exploited to further simplify the dynamics by applying another
adiabatic approximation, in which the phase difference $\delta_{2}-\delta_{1}$
is treated as a 'fast' variable and the phase average $\delta_{+}=\left(
\delta_{1}+\delta_{2}\right)  /2$ as a 'slow' one. To lowest order in the
adiabatic expansion one finds that for low frequencies $\omega\ll
\omega_{\mathrm{pl}}$, namely in the region where the impedance associated
with the capacitance of the JJs is much larger in absolute value in comparison
with the impedance associated with the inductance, the SQUID behaves as a
single JJ having critical current given by \cite{Tesche_380}%
\begin{equation}
I_{\mathrm{S}}=2I_{\mathrm{c}}\sqrt{1-\left(  1-\alpha^{2}\right)  \sin
^{2}\left(  \pi\Phi_{x}/\Phi_{0}\right)  }\;.
\end{equation}
Note that this approximation may break down when $\Phi_{x}\simeq\left(
n+1/2\right)  \Phi_{0}$ unless the asymmetry parameter $\alpha$ is
sufficiently large. The relatively large value of $\alpha$ in our device
($\alpha\simeq0.5$) ensures the validity of the above approximation. Using
this result, it is straightforward to obtain the following analytical approximations:%

\begin{subequations}
\begin{align}
\frac{\partial^{2}\left(  \varepsilon_{0}/E_{0}\right)  }{\partial\left(
I/I_{\mathrm{c}}\right)  ^{2}}  &  =\frac{I_{\mathrm{c}}}{\pi I_{\mathrm{S}}%
}\;,\\
\frac{\partial^{4}\left(  \varepsilon_{0}/E_{0}\right)  }{\partial\left(
I/I_{\mathrm{c}}\right)  ^{4}}  &  =-\frac{8}{3\pi^{2}}\left(  \frac
{I_{\mathrm{c}}}{I_{\mathrm{S}}}\right)  ^{3}\;,
\end{align}
which can be used to evaluate all the terms in Eq. (\ref{H_eff}).

\section{Appendix B: Hysteretic response and heating of the nanobridges}

As we discuss in the paper, the resonator exhibits hysteretic response to
magnetic flux when the input power is relatively low. Such a behavior occurs,
as can be seen from Eq. (\ref{u=}) above, when the screening parameter
$\beta_{L}$ is sufficiently large to give rise to metastability in the
dimensionless potential $u$. A fitting of the model to the experimental data
shown in Fig. 3(a) of the paper yields a value of $\beta_{L}=7.4$. Another
example of hysteretic response is shown in Fig. \ref{largebetal} below that
shows data taken with another sample, which was fabricated using the same
process that is described in the first section. The larger critical current in
that sample yields a larger value of the screening parameter $\beta_{L}=20$.%

\begin{figure}
[b]
\begin{center}
\includegraphics[
trim=0.000000in 0.000000in -0.013605in 0.000000in,
height=2.4241in,
width=3.2335in
]%
{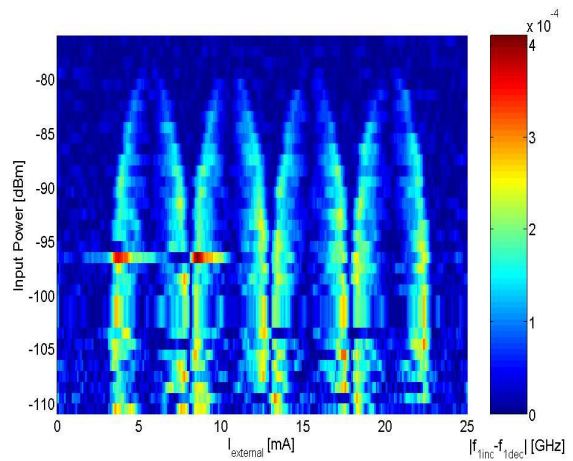}%
\caption{The difference between the measured resonance frequencies obtained in
the increasing flux sweep ($f_{\mathrm{1inc}})$ and the decreasing flux sweep
($f_{\mathrm{1dec}})$ of the first (\textit{detector}) mode. The dark blue
areas correspond to monostable regions, namely, the same resonance frequency
is measured for both the increased and decreased sweep. The red indicates the
regions where the system is bistable.}%
\label{inc_dec_dif}%
\end{center}
\end{figure}

As is mentioned in the paper, as the input power is increased the response
becomes non-hysteretic. The gradual transition between the hysteretic region
to the non-hysteretic one is seen in Fig. \ref{inc_dec_dif} below, which shows
the difference in the measured resonance frequency of the first mode obtained
from increased flux sweep $\left(  f_{\mathrm{1inc}}\right)  $ and decreased
flux sweep $\left(  f_{\mathrm{1dec}}\right)  $ at different input powers.
Dark blue in the color map corresponds to no difference, namely to monostable
regions, whereas in the red regions, where a large difference is observed, the
system is bistable. As can be clearly seen from the figure, the bistable
regions shrink as the input power is increased. The experimental results
suggest that the critical current of the nanobridges drops as the input power
is increased, and consequently the response becomes non-hysteretic due to the
resultant smaller value of the screening parameter $\beta_{L}$. We hypothesize
that the drop in the critical current occurs due to heating of the nanobridges
by the input power.

To estimate the effect of heating, we assume the case where the substrate is
isothermal and that the heat is mainly dissipated down into the substrate
rather than along the film \cite{Johnson_7069}. Moreover, we assume that most
of the externally injected power into the resonator is dissipated near the
nanobridges, where, the current density obtains its largest value. By
estimating the heat transfer coefficient per unit area between each nanobridge
and the substrate beneath it ($100%
\operatorname{nm}%
$ SiN on top of high-resistivity Si) to be $\kappa\simeq1%
\operatorname{W}%
\operatorname{cm}%
^{-2}%
\operatorname{K}%
^{-1}$ \cite{Monticone_3866,Weiser_4888} and the area of the nanobridge to be
$A\simeq\left(  50%
\operatorname{nm}%
\right)  ^{2}$one finds that the expected temperature rise for $P_{\mathrm{in}%
}=-70$ dBm is $\Delta T=P_{\mathrm{in}}/A\kappa\simeq4%
\operatorname{K}%
$.

Since heating is produced by AC current flowing through the nanobridges, it is
important to estimate also the thermal rate, which characterizes the inverse
of the typical time scale of thermalization, and is given by $\gamma
_{\mathrm{T}}=A\kappa/C$, where the heat capacity $C$ of the nanobridge is
given by $C=C_{v}Ad$, $C_{v}$ is the heat capacity per unit volume, and $d$ is
the thickness of the superconducting film. Using the estimate $C_{v}%
\simeq10^{-3}%
\operatorname{J}%
\operatorname{cm}%
^{-3}%
\operatorname{K}%
^{-1}$ \cite{Weiser_4888} one finds $\gamma_{\mathrm{T}}\simeq0.1%
\operatorname{GHz}%
$. Since the frequency of the AC heating current is 1-2 orders of magnitude
higher, we conclude that to a good approximation the temperature of the
nanobridges can be considered as stationary in the steady state.

\bibliographystyle{apsrev}
\bibliography{Articles,bib,Eyal_Bib,nanosensors}

\begin{thebibliography}{44}
\expandafter\ifx\csname natexlab\endcsname\relax\def\natexlab#1{#1}\fi
\expandafter\ifx\csname bibnamefont\endcsname\relax
  \def\bibnamefont#1{#1}\fi
\expandafter\ifx\csname bibfnamefont\endcsname\relax
  \def\bibfnamefont#1{#1}\fi
\expandafter\ifx\csname citenamefont\endcsname\relax
  \def\citenamefont#1{#1}\fi
\expandafter\ifx\csname url\endcsname\relax
  \def\url#1{\texttt{#1}}\fi
\expandafter\ifx\csname urlprefix\endcsname\relax\def\urlprefix{URL }\fi
\providecommand{\bibinfo}[2]{#2}
\providecommand{\eprint}[2][]{\url{#2}}

\bibitem[{\citenamefont{Zurek}(1991)}]{Zurek_36}
\bibinfo{author}{\bibfnamefont{W.~H.} \bibnamefont{Zurek}},
  \bibinfo{journal}{Physics Today} \textbf{\bibinfo{volume}{44}},
  \bibinfo{pages}{36} (\bibinfo{year}{1991}).

\bibitem[{\citenamefont{Imoto et~al.}(1985)\citenamefont{Imoto, Haus, and
  Yamamoto}}]{Imoto_2287}
\bibinfo{author}{\bibfnamefont{N.}~\bibnamefont{Imoto}},
  \bibinfo{author}{\bibfnamefont{H.~A.} \bibnamefont{Haus}}, \bibnamefont{and}
  \bibinfo{author}{\bibfnamefont{Y.}~\bibnamefont{Yamamoto}},
  \bibinfo{journal}{Phys. Rev. A} \textbf{\bibinfo{volume}{32}},
  \bibinfo{pages}{2287} (\bibinfo{year}{1985}).

\bibitem[{\citenamefont{Sanders and Milburn}(1989)}]{Sanders_694}
\bibinfo{author}{\bibfnamefont{B.~C.} \bibnamefont{Sanders}} \bibnamefont{and}
  \bibinfo{author}{\bibfnamefont{G.~J.} \bibnamefont{Milburn}},
  \bibinfo{journal}{Phys. Rev. A} \textbf{\bibinfo{volume}{39}},
  \bibinfo{pages}{694} (\bibinfo{year}{1989}).

\bibitem[{\citenamefont{Munro et~al.}(2005)\citenamefont{Munro, Nemoto,
  Beausoleil, and Spiller}}]{Munro_033819}
\bibinfo{author}{\bibfnamefont{W.~J.} \bibnamefont{Munro}},
  \bibinfo{author}{\bibfnamefont{K.}~\bibnamefont{Nemoto}},
  \bibinfo{author}{\bibfnamefont{R.~G.} \bibnamefont{Beausoleil}},
  \bibnamefont{and} \bibinfo{author}{\bibfnamefont{T.~P.}
  \bibnamefont{Spiller}}, \bibinfo{journal}{Physical Review A (Atomic,
  Molecular, and Optical Physics)} \textbf{\bibinfo{volume}{71}},
  \bibinfo{eid}{033819} (pages~\bibinfo{numpages}{4}) (\bibinfo{year}{2005}),
  \urlprefix\url{http://link.aps.org/abstract/PRA/v71/e033819}.

\bibitem[{\citenamefont{Santamore et~al.}(2004)\citenamefont{Santamore,
  Doherty, and Cross}}]{Santamore_144301}
\bibinfo{author}{\bibfnamefont{D.~H.} \bibnamefont{Santamore}},
  \bibinfo{author}{\bibfnamefont{A.~C.} \bibnamefont{Doherty}},
  \bibnamefont{and} \bibinfo{author}{\bibfnamefont{M.~C.} \bibnamefont{Cross}},
  \bibinfo{journal}{Phys. Rev. B} \textbf{\bibinfo{volume}{70}},
  \bibinfo{pages}{144301} (\bibinfo{year}{2004}).

\bibitem[{\citenamefont{Buks et~al.}(2008)\citenamefont{Buks, Segev, Zaitsev,
  Abdo, and Blencowe}}]{Buks_10001}
\bibinfo{author}{\bibfnamefont{E.}~\bibnamefont{Buks}},
  \bibinfo{author}{\bibfnamefont{E.}~\bibnamefont{Segev}},
  \bibinfo{author}{\bibfnamefont{S.}~\bibnamefont{Zaitsev}},
  \bibinfo{author}{\bibfnamefont{B.}~\bibnamefont{Abdo}}, \bibnamefont{and}
  \bibinfo{author}{\bibfnamefont{M.~P.} \bibnamefont{Blencowe}},
  \bibinfo{journal}{EPL} \textbf{\bibinfo{volume}{81}}, \bibinfo{pages}{10001}
  (\bibinfo{year}{2008}),
  \urlprefix\url{http://dx.doi.org/10.1209/0295-5075/81/10001}.

\bibitem[{\citenamefont{Helmer et~al.}(2008)\citenamefont{Helmer, Mariantoni,
  Solano, and Marquardt}}]{Helmer_0712_1908}
\bibinfo{author}{\bibfnamefont{F.}~\bibnamefont{Helmer}},
  \bibinfo{author}{\bibfnamefont{M.}~\bibnamefont{Mariantoni}},
  \bibinfo{author}{\bibfnamefont{E.}~\bibnamefont{Solano}}, \bibnamefont{and}
  \bibinfo{author}{\bibfnamefont{F.}~\bibnamefont{Marquardt}},
  \bibinfo{journal}{arXiv:0712.1908}  (\bibinfo{year}{2008}).

\bibitem[{\citenamefont{Buks and Yurke}(2006)}]{Buks_023815}
\bibinfo{author}{\bibfnamefont{E.}~\bibnamefont{Buks}} \bibnamefont{and}
  \bibinfo{author}{\bibfnamefont{B.}~\bibnamefont{Yurke}},
  \bibinfo{journal}{Phys. Rev. A} \textbf{\bibinfo{volume}{73}},
  \bibinfo{pages}{23815} (\bibinfo{year}{2006}).

\bibitem[{\citenamefont{Yamamoto et~al.}(2008)\citenamefont{Yamamoto, Inomata,
  Watanabe, Matsuba, Miyazaki, Oliver, Nakamura, and Tsai}}]{Yamamoto_042510}
\bibinfo{author}{\bibfnamefont{T.}~\bibnamefont{Yamamoto}},
  \bibinfo{author}{\bibfnamefont{K.}~\bibnamefont{Inomata}},
  \bibinfo{author}{\bibfnamefont{M.}~\bibnamefont{Watanabe}},
  \bibinfo{author}{\bibfnamefont{K.}~\bibnamefont{Matsuba}},
  \bibinfo{author}{\bibfnamefont{T.}~\bibnamefont{Miyazaki}},
  \bibinfo{author}{\bibfnamefont{W.~D.} \bibnamefont{Oliver}},
  \bibinfo{author}{\bibfnamefont{Y.}~\bibnamefont{Nakamura}}, \bibnamefont{and}
  \bibinfo{author}{\bibfnamefont{J.~S.} \bibnamefont{Tsai}},
  \bibinfo{journal}{Appl. Phys. Lett.} \textbf{\bibinfo{volume}{93}},
  \bibinfo{pages}{42510} (\bibinfo{year}{2008}).

\bibitem[{\citenamefont{Sandberg et~al.}(2008)\citenamefont{Sandberg, Wilson,
  Persson, Johansson, Shumeiko, Delsing, and Duty}}]{Sandberg_203501}
\bibinfo{author}{\bibfnamefont{M.}~\bibnamefont{Sandberg}},
  \bibinfo{author}{\bibfnamefont{C.~M.} \bibnamefont{Wilson}},
  \bibinfo{author}{\bibfnamefont{F.}~\bibnamefont{Persson}},
  \bibinfo{author}{\bibfnamefont{G.}~\bibnamefont{Johansson}},
  \bibinfo{author}{\bibfnamefont{V.}~\bibnamefont{Shumeiko}},
  \bibinfo{author}{\bibfnamefont{P.}~\bibnamefont{Delsing}}, \bibnamefont{and}
  \bibinfo{author}{\bibfnamefont{T.}~\bibnamefont{Duty}},
  \bibinfo{journal}{Appl. Phys. Lett.} \textbf{\bibinfo{volume}{92}},
  \bibinfo{pages}{203501} (\bibinfo{year}{2008}).

\bibitem[{\citenamefont{Palacios-Laloy
  et~al.}(2008)\citenamefont{Palacios-Laloy, Nguyen, Mallet, Bertet, Vion, and
  Esteve}}]{Palacios-Laloy_1034}
\bibinfo{author}{\bibfnamefont{A.}~\bibnamefont{Palacios-Laloy}},
  \bibinfo{author}{\bibfnamefont{F.}~\bibnamefont{Nguyen}},
  \bibinfo{author}{\bibfnamefont{F.}~\bibnamefont{Mallet}},
  \bibinfo{author}{\bibfnamefont{P.}~\bibnamefont{Bertet}},
  \bibinfo{author}{\bibfnamefont{D.}~\bibnamefont{Vion}}, \bibnamefont{and}
  \bibinfo{author}{\bibfnamefont{D.}~\bibnamefont{Esteve}},
  \bibinfo{journal}{Journal of Low Temperature Physics}
  \textbf{\bibinfo{volume}{151}}, \bibinfo{pages}{1034} (\bibinfo{year}{2008}),
  \urlprefix\url{http://dx.doi.org/10.1007/s10909-008-9774-x}.

\bibitem[{\citenamefont{Clark et~al.}(2001)\citenamefont{Clark, Prance,
  Whiteman, Prance, Everitt, Bulsara, and Ralph}}]{Clark_3042}
\bibinfo{author}{\bibfnamefont{T.~D.} \bibnamefont{Clark}},
  \bibinfo{author}{\bibfnamefont{R.~J.} \bibnamefont{Prance}},
  \bibinfo{author}{\bibfnamefont{R.}~\bibnamefont{Whiteman}},
  \bibinfo{author}{\bibfnamefont{H.}~\bibnamefont{Prance}},
  \bibinfo{author}{\bibfnamefont{M.~J.} \bibnamefont{Everitt}},
  \bibinfo{author}{\bibfnamefont{A.~R.} \bibnamefont{Bulsara}},
  \bibnamefont{and} \bibinfo{author}{\bibfnamefont{J.~F.} \bibnamefont{Ralph}},
  \bibinfo{journal}{Journal of Applied Physics} \textbf{\bibinfo{volume}{90}},
  \bibinfo{pages}{3042} (\bibinfo{year}{2001}),
  \urlprefix\url{http://link.aip.org/link/?JAP/90/3042/1}.

\bibitem[{\citenamefont{Nation et~al.}(2008)\citenamefont{Nation, Blencowe, and
  Buks}}]{Nation_104516}
\bibinfo{author}{\bibfnamefont{P.~D.} \bibnamefont{Nation}},
  \bibinfo{author}{\bibfnamefont{M.~P.} \bibnamefont{Blencowe}},
  \bibnamefont{and} \bibinfo{author}{\bibfnamefont{E.}~\bibnamefont{Buks}},
  \bibinfo{journal}{Phys. Rev. B} \textbf{\bibinfo{volume}{78}},
  \bibinfo{pages}{104516} (\bibinfo{year}{2008}).

\bibitem[{\citenamefont{Likharev}(1979)}]{Likharev_101}
\bibinfo{author}{\bibfnamefont{K.~K.} \bibnamefont{Likharev}},
  \bibinfo{journal}{Rev. Mod. Phys.} \textbf{\bibinfo{volume}{51}},
  \bibinfo{pages}{101} (\bibinfo{year}{1979}).

\bibitem[{\citenamefont{Troeman et~al.}(2008)\citenamefont{Troeman, van~der
  Ploeg, Il'Ichev, Meyer, Golubov, Kupriyanov, and
  Hilgenkamp}}]{Troeman_024509}
\bibinfo{author}{\bibfnamefont{A.~G.~P.} \bibnamefont{Troeman}},
  \bibinfo{author}{\bibfnamefont{S.~H.~W.} \bibnamefont{van~der Ploeg}},
  \bibinfo{author}{\bibfnamefont{E.}~\bibnamefont{Il'Ichev}},
  \bibinfo{author}{\bibfnamefont{H.-G.} \bibnamefont{Meyer}},
  \bibinfo{author}{\bibfnamefont{A.~A.} \bibnamefont{Golubov}},
  \bibinfo{author}{\bibfnamefont{M.~Y.} \bibnamefont{Kupriyanov}},
  \bibnamefont{and}
  \bibinfo{author}{\bibfnamefont{H.}~\bibnamefont{Hilgenkamp}},
  \bibinfo{journal}{Physical Review B (Condensed Matter and Materials Physics)}
  \textbf{\bibinfo{volume}{77}}, \bibinfo{eid}{024509}
  (pages~\bibinfo{numpages}{5}) (\bibinfo{year}{2008}),
  \urlprefix\url{http://link.aps.org/abstract/PRB/v77/e024509}.

\bibitem[{\citenamefont{Lam and Tilbrook}(2003)}]{Lam_1078}
\bibinfo{author}{\bibfnamefont{S.~K.~H.} \bibnamefont{Lam}} \bibnamefont{and}
  \bibinfo{author}{\bibfnamefont{D.~L.} \bibnamefont{Tilbrook}},
  \bibinfo{journal}{Applied Physics Letters} \textbf{\bibinfo{volume}{82}},
  \bibinfo{pages}{1078} (\bibinfo{year}{2003}),
  \urlprefix\url{http://link.aip.org/link/?APL/82/1078/1}.

\bibitem[{\citenamefont{Podd et~al.}(2007)\citenamefont{Podd, Hutchinson,
  Williams, and Hasko}}]{Podd_134501}
\bibinfo{author}{\bibfnamefont{G.~J.} \bibnamefont{Podd}},
  \bibinfo{author}{\bibfnamefont{G.~D.} \bibnamefont{Hutchinson}},
  \bibinfo{author}{\bibfnamefont{D.~A.} \bibnamefont{Williams}},
  \bibnamefont{and} \bibinfo{author}{\bibfnamefont{D.~G.} \bibnamefont{Hasko}},
  \bibinfo{journal}{Physical Review B (Condensed Matter and Materials Physics)}
  \textbf{\bibinfo{volume}{75}}, \bibinfo{pages}{134501}
  (\bibinfo{year}{2007}).

\bibitem[{\citenamefont{Hao et~al.}(2008)\citenamefont{Hao, Macfarlane, Gallop,
  Cox, Beyer, Drung, and Schurig}}]{Hao_192507}
\bibinfo{author}{\bibfnamefont{L.}~\bibnamefont{Hao}},
  \bibinfo{author}{\bibfnamefont{J.~C.} \bibnamefont{Macfarlane}},
  \bibinfo{author}{\bibfnamefont{J.~C.} \bibnamefont{Gallop}},
  \bibinfo{author}{\bibfnamefont{D.}~\bibnamefont{Cox}},
  \bibinfo{author}{\bibfnamefont{J.}~\bibnamefont{Beyer}},
  \bibinfo{author}{\bibfnamefont{D.}~\bibnamefont{Drung}}, \bibnamefont{and}
  \bibinfo{author}{\bibfnamefont{T.}~\bibnamefont{Schurig}},
  \bibinfo{journal}{Applied Physics Letters} \textbf{\bibinfo{volume}{92}},
  \bibinfo{eid}{192507} (pages~\bibinfo{numpages}{3}) (\bibinfo{year}{2008}),
  \urlprefix\url{http://link.aip.org/link/?APL/92/192507/1}.

\bibitem[{\citenamefont{Hao et~al.}(2007)\citenamefont{Hao, Macfarlane, Gallop,
  Cox, Joseph-Franks, Hutson, Chen, and Lam}}]{Hao_392}
\bibinfo{author}{\bibfnamefont{L.}~\bibnamefont{Hao}},
  \bibinfo{author}{\bibfnamefont{J.~C.} \bibnamefont{Macfarlane}},
  \bibinfo{author}{\bibfnamefont{J.~C.} \bibnamefont{Gallop}},
  \bibinfo{author}{\bibfnamefont{D.}~\bibnamefont{Cox}},
  \bibinfo{author}{\bibfnamefont{P.}~\bibnamefont{Joseph-Franks}},
  \bibinfo{author}{\bibfnamefont{D.}~\bibnamefont{Hutson}},
  \bibinfo{author}{\bibfnamefont{J.}~\bibnamefont{Chen}}, \bibnamefont{and}
  \bibinfo{author}{\bibfnamefont{S.~K.~H.} \bibnamefont{Lam}},
  \bibinfo{journal}{IEEE Transactions on Instrumentation and Measurement}
  \textbf{\bibinfo{volume}{56}}, \bibinfo{pages}{392} (\bibinfo{year}{2007}).

\bibitem[{\citenamefont{Bell et~al.}(2003)\citenamefont{Bell, Burnell, Kang,
  Hadfield, Kappers, and Blamire}}]{Bell_630}
\bibinfo{author}{\bibfnamefont{C.}~\bibnamefont{Bell}},
  \bibinfo{author}{\bibfnamefont{G.}~\bibnamefont{Burnell}},
  \bibinfo{author}{\bibfnamefont{D.-J.} \bibnamefont{Kang}},
  \bibinfo{author}{\bibfnamefont{R.~H.} \bibnamefont{Hadfield}},
  \bibinfo{author}{\bibfnamefont{M.~J.} \bibnamefont{Kappers}},
  \bibnamefont{and} \bibinfo{author}{\bibfnamefont{M.~G.}
  \bibnamefont{Blamire}}, \bibinfo{journal}{Nanotechnology}
  \textbf{\bibinfo{volume}{14}}, \bibinfo{pages}{630} (\bibinfo{year}{2003}),
  \urlprefix\url{http://stacks.iop.org/0957-4484/14/630}.

\bibitem[{\citenamefont{Datesman
  et~al.}(2005{\natexlab{a}})\citenamefont{Datesman, Schultz, Lichtenberger,
  Golish, Walker, and Kooi}}]{Datesman_928}
\bibinfo{author}{\bibfnamefont{A.}~\bibnamefont{Datesman}},
  \bibinfo{author}{\bibfnamefont{J.}~\bibnamefont{Schultz}},
  \bibinfo{author}{\bibfnamefont{A.}~\bibnamefont{Lichtenberger}},
  \bibinfo{author}{\bibfnamefont{D.}~\bibnamefont{Golish}},
  \bibinfo{author}{\bibfnamefont{C.}~\bibnamefont{Walker}}, \bibnamefont{and}
  \bibinfo{author}{\bibfnamefont{J.}~\bibnamefont{Kooi}},
  \bibinfo{journal}{IEEE Transactions on Applied Superconductivity}
  \textbf{\bibinfo{volume}{15}}, \bibinfo{pages}{928}
  (\bibinfo{year}{2005}{\natexlab{a}}).

\bibitem[{\citenamefont{Troeman et~al.}(2007)\citenamefont{Troeman, Derking,
  Borger, Pleikies, Veldhuis, and Hilgenkamp}}]{Troeman_2152}
\bibinfo{author}{\bibfnamefont{A.}~\bibnamefont{Troeman}},
  \bibinfo{author}{\bibfnamefont{H.}~\bibnamefont{Derking}},
  \bibinfo{author}{\bibfnamefont{B.}~\bibnamefont{Borger}},
  \bibinfo{author}{\bibfnamefont{J.}~\bibnamefont{Pleikies}},
  \bibinfo{author}{\bibfnamefont{D.}~\bibnamefont{Veldhuis}}, \bibnamefont{and}
  \bibinfo{author}{\bibfnamefont{H.}~\bibnamefont{Hilgenkamp}},
  \bibinfo{journal}{Nano Letters} \textbf{\bibinfo{volume}{7}},
  \bibinfo{pages}{2152} (\bibinfo{year}{2007}), ISSN \bibinfo{issn}{1530-6984},
  \urlprefix\url{http://pubs3.acs.org/acs/journals/doilookup?in_doi=10.1021/nl%
070870f}.

\bibitem[{\citenamefont{Datesman
  et~al.}(2005{\natexlab{b}})\citenamefont{Datesman, Schultz, Cecil, Lyons, and
  Lichtenberger}}]{Datesman_3524}
\bibinfo{author}{\bibfnamefont{A.}~\bibnamefont{Datesman}},
  \bibinfo{author}{\bibfnamefont{J.}~\bibnamefont{Schultz}},
  \bibinfo{author}{\bibfnamefont{T.}~\bibnamefont{Cecil}},
  \bibinfo{author}{\bibfnamefont{C.}~\bibnamefont{Lyons}}, \bibnamefont{and}
  \bibinfo{author}{\bibfnamefont{A.}~\bibnamefont{Lichtenberger}},
  \bibinfo{journal}{IEEE Transactions on Applied Superconductivity}
  \textbf{\bibinfo{volume}{15}}, \bibinfo{pages}{3524}
  (\bibinfo{year}{2005}{\natexlab{b}}), ISSN \bibinfo{issn}{1051-8223}.

\bibitem[{\citenamefont{Yurke and Buks}(2006)}]{Yurke_5054}
\bibinfo{author}{\bibfnamefont{B.}~\bibnamefont{Yurke}} \bibnamefont{and}
  \bibinfo{author}{\bibfnamefont{E.}~\bibnamefont{Buks}}, \bibinfo{journal}{J.
  Lightwave Tech.} \textbf{\bibinfo{volume}{24}}, \bibinfo{pages}{5054}
  (\bibinfo{year}{2006}).

\bibitem[{\citenamefont{Levinson}(1997)}]{Levinson_299}
\bibinfo{author}{\bibfnamefont{Y.}~\bibnamefont{Levinson}},
  \bibinfo{journal}{Europhys. Lett.} \textbf{\bibinfo{volume}{39}},
  \bibinfo{pages}{299} (\bibinfo{year}{1997}).

\bibitem[{\citenamefont{Blencowe and Buks}(2007)}]{Blencowe_014511}
\bibinfo{author}{\bibfnamefont{M.~P.} \bibnamefont{Blencowe}} \bibnamefont{and}
  \bibinfo{author}{\bibfnamefont{E.}~\bibnamefont{Buks}},
  \bibinfo{journal}{Phys. Rev. B} \textbf{\bibinfo{volume}{76}},
  \bibinfo{pages}{14511} (\bibinfo{year}{2007}).

\bibitem[{\citenamefont{Granata et~al.}(2008)\citenamefont{Granata, Esposito,
  Vettoliere, Petti, and Russo}}]{Granata_275501}
\bibinfo{author}{\bibfnamefont{C.}~\bibnamefont{Granata}},
  \bibinfo{author}{\bibfnamefont{E.}~\bibnamefont{Esposito}},
  \bibinfo{author}{\bibfnamefont{A.}~\bibnamefont{Vettoliere}},
  \bibinfo{author}{\bibfnamefont{L.}~\bibnamefont{Petti}}, \bibnamefont{and}
  \bibinfo{author}{\bibfnamefont{M.}~\bibnamefont{Russo}},
  \bibinfo{journal}{Nanotechnology} \textbf{\bibinfo{volume}{19}},
  \bibinfo{pages}{275501} (\bibinfo{year}{2008}).

\bibitem[{\citenamefont{Hasselbach et~al.}(2002)\citenamefont{Hasselbach,
  Mailly, and Kirtley}}]{Hasselbach_4432}
\bibinfo{author}{\bibfnamefont{K.}~\bibnamefont{Hasselbach}},
  \bibinfo{author}{\bibfnamefont{D.}~\bibnamefont{Mailly}}, \bibnamefont{and}
  \bibinfo{author}{\bibfnamefont{J.}~\bibnamefont{Kirtley}},
  \bibinfo{journal}{Journal of Applied Physics} \textbf{\bibinfo{volume}{91}},
  \bibinfo{pages}{4432} (\bibinfo{year}{2002}).

\bibitem[{\citenamefont{Hasselbach et~al.}(2000)\citenamefont{Hasselbach,
  Veauvy, and Mailly}}]{Hasselbach_140}
\bibinfo{author}{\bibfnamefont{K.}~\bibnamefont{Hasselbach}},
  \bibinfo{author}{\bibfnamefont{C.}~\bibnamefont{Veauvy}}, \bibnamefont{and}
  \bibinfo{author}{\bibfnamefont{D.}~\bibnamefont{Mailly}},
  \bibinfo{journal}{Physica C Superconductivity}
  \textbf{\bibinfo{volume}{332}}, \bibinfo{pages}{140} (\bibinfo{year}{2000}).

\bibitem[{\citenamefont{Baratoff et~al.}(1970)\citenamefont{Baratoff,
  Blackburn, and Schwartz}}]{Baratoff_1096}
\bibinfo{author}{\bibfnamefont{A.}~\bibnamefont{Baratoff}},
  \bibinfo{author}{\bibfnamefont{J.~A.} \bibnamefont{Blackburn}},
  \bibnamefont{and} \bibinfo{author}{\bibfnamefont{B.~B.}
  \bibnamefont{Schwartz}}, \bibinfo{journal}{Phys. Rev. Lett.}
  \textbf{\bibinfo{volume}{25}}, \bibinfo{pages}{1096} (\bibinfo{year}{1970}).

\bibitem[{\citenamefont{Likharev and Yakobson}(1975)}]{Likharev_950}
\bibinfo{author}{\bibfnamefont{K.~K.} \bibnamefont{Likharev}} \bibnamefont{and}
  \bibinfo{author}{\bibfnamefont{L.~A.} \bibnamefont{Yakobson}},
  \bibinfo{journal}{Sov. Phys. - Tech. Phys. (Engl. Transl.)}
  \textbf{\bibinfo{volume}{20}}, \bibinfo{pages}{950} (\bibinfo{year}{1975}).

\bibitem[{\citenamefont{Gumann et~al.}(2007)\citenamefont{Gumann, Dahm, and
  Schopohl}}]{Gumann_064529}
\bibinfo{author}{\bibfnamefont{A.}~\bibnamefont{Gumann}},
  \bibinfo{author}{\bibfnamefont{T.}~\bibnamefont{Dahm}}, \bibnamefont{and}
  \bibinfo{author}{\bibfnamefont{N.}~\bibnamefont{Schopohl}},
  \bibinfo{journal}{Physical Review B (Condensed Matter and Materials Physics)}
  \textbf{\bibinfo{volume}{76}}, \bibinfo{eid}{064529}
  (pages~\bibinfo{numpages}{14}) (\bibinfo{year}{2007}),
  \urlprefix\url{http://link.aps.org/abstract/PRB/v76/e064529}.

\bibitem[{\citenamefont{Pronin et~al.}(1998)\citenamefont{Pronin, Dressel,
  Pimenov, Loidl, Roshchin, and Greene}}]{Pronin_14416}
\bibinfo{author}{\bibfnamefont{A.~V.} \bibnamefont{Pronin}},
  \bibinfo{author}{\bibfnamefont{M.}~\bibnamefont{Dressel}},
  \bibinfo{author}{\bibfnamefont{A.}~\bibnamefont{Pimenov}},
  \bibinfo{author}{\bibfnamefont{A.}~\bibnamefont{Loidl}},
  \bibinfo{author}{\bibfnamefont{I.~V.} \bibnamefont{Roshchin}},
  \bibnamefont{and} \bibinfo{author}{\bibfnamefont{L.~H.}
  \bibnamefont{Greene}}, \bibinfo{journal}{Phys. Rev. B}
  \textbf{\bibinfo{volume}{57}}, \bibinfo{pages}{14416} (\bibinfo{year}{1998}).

\bibitem[{\citenamefont{Maxfield and McLean}(1965)}]{Maxfield_A1515}
\bibinfo{author}{\bibfnamefont{B.~W.} \bibnamefont{Maxfield}} \bibnamefont{and}
  \bibinfo{author}{\bibfnamefont{W.~L.} \bibnamefont{McLean}},
  \bibinfo{journal}{Phys. Rev.} \textbf{\bibinfo{volume}{139}},
  \bibinfo{pages}{A1515} (\bibinfo{year}{1965}).

\bibitem[{\citenamefont{Golubov et~al.}(2004)\citenamefont{Golubov, Kupriyanov,
  and Il\char39{}ichev}}]{Golubov_411}
\bibinfo{author}{\bibfnamefont{A.~A.} \bibnamefont{Golubov}},
  \bibinfo{author}{\bibfnamefont{M.~Y.} \bibnamefont{Kupriyanov}},
  \bibnamefont{and}
  \bibinfo{author}{\bibfnamefont{E.}~\bibnamefont{Il\char39{}ichev}},
  \bibinfo{journal}{Rev. Mod. Phys.} \textbf{\bibinfo{volume}{76}},
  \bibinfo{pages}{411} (\bibinfo{year}{2004}).

\bibitem[{\citenamefont{Mitra et~al.}(2008)\citenamefont{Mitra, Strauch, Lobb,
  Anderson, Wellstood, and Tiesinga}}]{Mitra_214512}
\bibinfo{author}{\bibfnamefont{K.}~\bibnamefont{Mitra}},
  \bibinfo{author}{\bibfnamefont{F.~W.} \bibnamefont{Strauch}},
  \bibinfo{author}{\bibfnamefont{C.~J.} \bibnamefont{Lobb}},
  \bibinfo{author}{\bibfnamefont{J.~R.} \bibnamefont{Anderson}},
  \bibinfo{author}{\bibfnamefont{F.~C.} \bibnamefont{Wellstood}},
  \bibnamefont{and} \bibinfo{author}{\bibfnamefont{E.}~\bibnamefont{Tiesinga}},
  \bibinfo{journal}{Physical Review B (Condensed Matter and Materials Physics)}
  \textbf{\bibinfo{volume}{77}}, \bibinfo{eid}{214512}
  (pages~\bibinfo{numpages}{10}) (\bibinfo{year}{2008}),
  \urlprefix\url{http://link.aps.org/abstract/PRB/v77/e214512}.

\bibitem[{\citenamefont{Ralph et~al.}(1996)\citenamefont{Ralph, Clark, Prance,
  Prance, and Diggins}}]{Ralph_10753}
\bibinfo{author}{\bibfnamefont{J.~F.} \bibnamefont{Ralph}},
  \bibinfo{author}{\bibfnamefont{T.~D.} \bibnamefont{Clark}},
  \bibinfo{author}{\bibfnamefont{R.~J.} \bibnamefont{Prance}},
  \bibinfo{author}{\bibfnamefont{H.}~\bibnamefont{Prance}}, \bibnamefont{and}
  \bibinfo{author}{\bibfnamefont{J.}~\bibnamefont{Diggins}},
  \bibinfo{journal}{J. Phys.: Condens. Matter} \textbf{\bibinfo{volume}{8}},
  \bibinfo{pages}{10753} (\bibinfo{year}{1996}).

\bibitem[{\citenamefont{Buks et~al.}(2007)\citenamefont{Buks, Zaitsev, Segev,
  Abdo, and Blencowe}}]{Buks_026217}
\bibinfo{author}{\bibfnamefont{E.}~\bibnamefont{Buks}},
  \bibinfo{author}{\bibfnamefont{S.}~\bibnamefont{Zaitsev}},
  \bibinfo{author}{\bibfnamefont{E.}~\bibnamefont{Segev}},
  \bibinfo{author}{\bibfnamefont{B.}~\bibnamefont{Abdo}}, \bibnamefont{and}
  \bibinfo{author}{\bibfnamefont{M.~P.} \bibnamefont{Blencowe}},
  \bibinfo{journal}{Phys. Rev. E} \textbf{\bibinfo{volume}{76}},
  \bibinfo{pages}{26217} (\bibinfo{year}{2007}).

\bibitem[{\citenamefont{Littlejohn and Flynn}(1991)}]{Littlejohn_5239}
\bibinfo{author}{\bibfnamefont{R.~G.} \bibnamefont{Littlejohn}}
  \bibnamefont{and} \bibinfo{author}{\bibfnamefont{W.~G.} \bibnamefont{Flynn}},
  \bibinfo{journal}{Phys. Rev. A} \textbf{\bibinfo{volume}{44}},
  \bibinfo{pages}{5239} (\bibinfo{year}{1991}).

\bibitem[{\citenamefont{Panati et~al.}(2002)\citenamefont{Panati, Spohn, and
  Teufel}}]{Panati_250405}
\bibinfo{author}{\bibfnamefont{G.}~\bibnamefont{Panati}},
  \bibinfo{author}{\bibfnamefont{H.}~\bibnamefont{Spohn}}, \bibnamefont{and}
  \bibinfo{author}{\bibfnamefont{S.}~\bibnamefont{Teufel}},
  \bibinfo{journal}{Phys. Rev. Lett.} \textbf{\bibinfo{volume}{88}},
  \bibinfo{pages}{250405} (\bibinfo{year}{2002}).

\bibitem[{\citenamefont{Tesche and Clarke}(1977)}]{Tesche_380}
\bibinfo{author}{\bibfnamefont{C.~D.} \bibnamefont{Tesche}} \bibnamefont{and}
  \bibinfo{author}{\bibfnamefont{J.}~\bibnamefont{Clarke}},
  \bibinfo{journal}{J. low Temp. Phys.} \textbf{\bibinfo{volume}{29}},
  \bibinfo{pages}{301} (\bibinfo{year}{1977}).

\bibitem[{\citenamefont{Johnson et~al.}(1996)\citenamefont{Johnson, Herr, and
  Kadin}}]{Johnson_7069}
\bibinfo{author}{\bibfnamefont{M.~W.} \bibnamefont{Johnson}},
  \bibinfo{author}{\bibfnamefont{A.~M.} \bibnamefont{Herr}}, \bibnamefont{and}
  \bibinfo{author}{\bibfnamefont{A.~M.} \bibnamefont{Kadin}},
  \bibinfo{journal}{J. Appl. Phys.} \textbf{\bibinfo{volume}{79}},
  \bibinfo{pages}{7069} (\bibinfo{year}{1996}).

\bibitem[{\citenamefont{Monticone et~al.}(1999)\citenamefont{Monticone,
  Lacquaniti, Steni, Rajteri, Rastello, and Parlato}}]{Monticone_3866}
\bibinfo{author}{\bibfnamefont{E.}~\bibnamefont{Monticone}},
  \bibinfo{author}{\bibfnamefont{V.}~\bibnamefont{Lacquaniti}},
  \bibinfo{author}{\bibfnamefont{R.}~\bibnamefont{Steni}},
  \bibinfo{author}{\bibfnamefont{M.}~\bibnamefont{Rajteri}},
  \bibinfo{author}{\bibfnamefont{M.}~\bibnamefont{Rastello}}, \bibnamefont{and}
  \bibinfo{author}{\bibfnamefont{L.}~\bibnamefont{Parlato}},
  \bibinfo{journal}{IEEE Trans. Appl. Super.} \textbf{\bibinfo{volume}{9}},
  \bibinfo{pages}{3866} (\bibinfo{year}{1999}).

\bibitem[{\citenamefont{Weiser et~al.}(1981)\citenamefont{Weiser, Strom, Wolf,
  and Gubser}}]{Weiser_4888}
\bibinfo{author}{\bibfnamefont{K.}~\bibnamefont{Weiser}},
  \bibinfo{author}{\bibfnamefont{U.}~\bibnamefont{Strom}},
  \bibinfo{author}{\bibfnamefont{S.~A.} \bibnamefont{Wolf}}, \bibnamefont{and}
  \bibinfo{author}{\bibfnamefont{D.~U.} \bibnamefont{Gubser}},
  \bibinfo{journal}{J. Appl. Phys.} \textbf{\bibinfo{volume}{52}},
  \bibinfo{pages}{4888} (\bibinfo{year}{1981}).

\end{thebibliography}

\end{subequations}
\end{document}